\documentclass[twocolumn,showpacs,preprintnumbers,amsmath,amssymb]{revtex4}

\usepackage{graphicx}
\usepackage{dcolumn}
\usepackage{bm}

\begin{document}

\preprint{APS/123-QED}

\title{Collective excitations of a trapped Bose-Einstein condensate
in the presence of a 1D optical lattice }

\author {C.~Fort, F.~S.~Cataliotti, L.~Fallani, F.~Ferlaino, P.~Maddaloni, and M.~Inguscio}
\affiliation{LENS, Dipartimento di Fisica, Universit\`a di Firenze
and Istituto Nazionale per la Fisica della Materia, via Nello
Carrara 1, I-50019 Sesto Fiorentino (Firenze), Italy}

\date{\today}

\begin{abstract}
We study low-lying collective modes of a horizontally elongated
$^{87}$Rb condensate produced in a 3D magnetic harmonic trap with
the addition of a 1D periodic potential which is provided by a
laser standing-wave along the horizontal axis. While the
transverse breathing mode results unperturbed, quadrupole and
dipole oscillations along the optical lattice are strongly
modified. Precise measurements of the collective mode frequencies
at different height of the optical barriers provide a stringent
test of the theoretical model recently introduced [M.~Kr\"{a}mer
{\it et al.} Phys. Rev. Lett. {\bf 88} 180404 (2002)].

\end{abstract}

 \narrowtext

\maketitle

The measurement of frequencies of collective modes was immediately
recognized to be a fundamental and precise tool to investigate the
quantum macroscopic behaviour of atomic Bose-Einstein condensates.
The observations of low-lying excitations \cite{JILA,MIT,noi},
including the scissor mode \cite{Foot}, were an important step
toward the characterization of these systems. Collective
excitations were also studied at finite temperature
\cite{JILAT,MITT,FootT,dalibard} investigating frequency shifts
and damping rates. All the experiments so far reported were
performed on Bose-Einstein condensates magnetically trapped in
pure harmonic potentials.

More recently, condensates trapped in periodic potentials have
attracted much interest demonstrated by the flourishing of both
theoretical and experimental papers in this field. BEC trapped in
a periodic potential is particularly interesting as a possible
system for the investigation of pure quantum effects. Many results
have been reported e.g. interference of atomic de Broglie waves
tunneling from a vertical array of BECs \cite{kasevich1},
observation of squeezed states in a BEC \cite{kasevich2},
realization of a linear array of Josephson junctions
\cite{science} with BECs, observation of Bloch oscillations
\cite{arimondoPRLbloch} and culminating with the observation of
the quantum phase transition from a superfluid to a Mott insulator
\cite{mott}. Recently, the general interest of this field has been
extended to a careful study of the loading of BECs in an optical
lattice \cite{rolston} and to the observation of collapse and
revival of the matter wave field of a BEC \cite{greiner02}. In
this context, the characterization of collective modes of a
condensate in the presence of an optical lattice motivated the
development of a new theoretical treatment \cite{kraemer} based on
a generalization of the hydrodynamic equation of superfluids for a
weakly interacting Bose gas \cite{stringariPRL96,dalfovoRMP} to
include the effects of the periodic potential. The collective
modes of a trapped BEC are predicted to be significantly modified
by the presence of an optical lattice. Precise measurements of the
low-lying collective frequency would verify the validity of the
mass renormalization theory \cite{kraemer}.

In this Letter we quantitatively investigate the modification of
the low-lying excitation spectrum of a harmonically trapped BEC,
due to the presence of a 1D optical lattice. In particular we
measure the frequencies of the quadrupole and transverse breathing
modes as a function of the optical lattice depth. The experimental
observations are compared with the predictions of the model
reported in \cite{kraemer} consisting in a frequency shift of the
collective modes characterized by a motion along the lattice axis,
whereas the frequency of collective modes involving an atomic flow
transverse respect to the periodic potential, is expected to
remain unchanged.

The experiment is performed with a $F=1, m_F=-1$ $^{87}$Rb
condensate produced in the combined potential obtained
superimposing a 1D optical lattice along the axial direction of a
Ioffe-Pritchard magnetic trap \cite{science,pedri}. The harmonic
magnetic trap is characterized by a cylindrical symmetry with
axial and radial frequencies respectively $\nu_z=8.70\pm 0.02$~Hz
and $\nu_{\perp}=85.7\pm 0.6$~Hz. The 1D optical periodic
potential is provided by retroreflecting along the axial direction
a far detuned laser beam. The light is produced by a commercial
Ti:Sapphire laser tuned at $\lambda=757$~nm. The radial dimension
of the laser beam in correspondence of the condensate is $\sim
300$~$\mu$m, large enough, compared to the radial size of the
condensate, to neglect the effect of the optical potential on the
radial dynamics. The optical potential height $V_{opt}$ has been
varied in the experiment up to 5.2~$E_r$, $E_r=h^2/2m \lambda^2$
being the recoil energy of an atom of mass $m$ absorbing one
lattice photon. The corresponding photon scattering rate, below
0.04~s$^{-1}$, is negligible in the time scale of our experiment.
In the experiment the lattice depth spans from the weak- to the
tight-binding regime in order to investigate the generality of the
theoretical predictions in \cite{kraemer}.

After the condensate has been produced, in order to excite the
collective modes, we perturb the magnetic bias field
\cite{JILA,MIT,noi} by applying five cycles of resonant sinusoidal
modulation, thus producing a periodical perturbation of the radial
frequency of the magnetic trap. We modulate the radial frequency
by 10\% of its value. A bigger oscillation amplitude would produce
a loss of superfluidity caused by entering regions of unstable
dynamics \cite{instabilita}. This procedure excites modes with
zero angular momentum along the symmetry axis of the trap. For an
elliptical trap, the two lowest energy modes of this type
correspond to in-phase oscillations of the width along $x$ and $y$
and out-of-phase along $z$ (quadrupole mode), and an in-phase
compressional mode along all directions (breathing mode). In the
Thomas-Fermi regime, for small amplitude oscillations and strongly
elongated traps, the two modes are characterized respectively by
the frequencies $\sqrt{5/2} \nu_z $ and $2 \nu_{\perp}$
\cite{stringariPRL96}. In this limit the two frequencies are quite
different and the axial and radial excitations are almost
decoupled. The axial width performs small amplitude oscillations
at $2 \nu_{\perp}$ superimposed to wider amplitude oscillations at
frequency $\sqrt{5/2} \nu_z $ (quadrupole mode), and vice versa
for the radial width (transverse breathing mode).

After exciting the collective mode, we let the cloud evolve for a
variable time $t$, then we switch off the combined trap, let the
cloud expand for 29~ms and take an absorption image of the
expanded cloud along one of the radial directions. In the regime
of large optical lattice heights the density profile after the
expansion results in an interferogram consisting of a central
cloud and two lateral peaks \cite{pedri}. From the image we
extract the radial ($R_{\perp}$) and axial ($R_z$) radii of the
central peak as a function of time $t$. In order to compensate the
effect of fluctuations in the condensates number of atoms we then
plot the aspect ratio $R_{\perp}/R_z$ and fit the data to obtain
the mode frequency.

{\it Quadrupole Mode}. Following \cite{kraemer}, a trapped BEC in
the presence of a 1D optical lattice can still be described by the
hydrodynamic equations that take the same form as in the absence
of the lattice once defined a renormalized interaction coupling
constant and an effective mass $m^*$. The effective mass depends
on the tunneling rate between adjacent optical wells thus
accounting for the modified inertia of the system along the
lattice. In particular, in the linear regime of small amplitude
oscillations, the new frequency of the collective modes is simply
obtained by replacing the axial magnetic trap frequency $\nu_z$
with $\nu_z \sqrt{m/m^*}$. In our experimental configuration,
where we have an elongated magnetic trap ($\nu_z \ll
\nu_{\perp}$), the dipole mode along the periodic potential and
the quadrupole mode are both shifted and characterized
respectively by the frequencies
\begin{eqnarray}
\label{frequenze}
&&\nu_D=\sqrt{\frac{m}{m^*}}\nu_z\\
\nonumber &&\nu_Q=\sqrt{\frac{5}{2}}\sqrt{\frac{m}{m^*}}\nu_z=
\sqrt{\frac{5}{2}} \nu_D
\end{eqnarray}
To resonantly excite the quadrupole mode we thus need to first
measure $\nu_D$. The measurement of the dipole mode frequency is
done as in \cite{science} where we already investigated the
dependence of the frequency of the dipole mode in conjunction with
current/phase dynamics in an array of Josephson junctions. In
particular, we induce dipole oscillations, by suddenly displacing
the position of the magnetic field minimum along $z$ and observing
the center-of-mass motion as a function of time. The quadrupole
mode is then excited by modulating the magnetic bias field at a
frequency close to $\sqrt{5/2} \nu_D$. The procedure to excite the
quadrupole mode takes $\sim$700~ms producing a significant heating
of the condensate. The final observed temperature of $\sim 150$~nK
($0.9$~$T_c$) is consistent with the heating rate of $64\pm7$~nK/s
measured in the absence of the optical lattice. The heating
results in a degradation of the interference pattern visibility so
that typically only the central peak is observable also for our
larger laser intensities.

\begin{figure}
\begin{center}
\includegraphics[width=8cm]{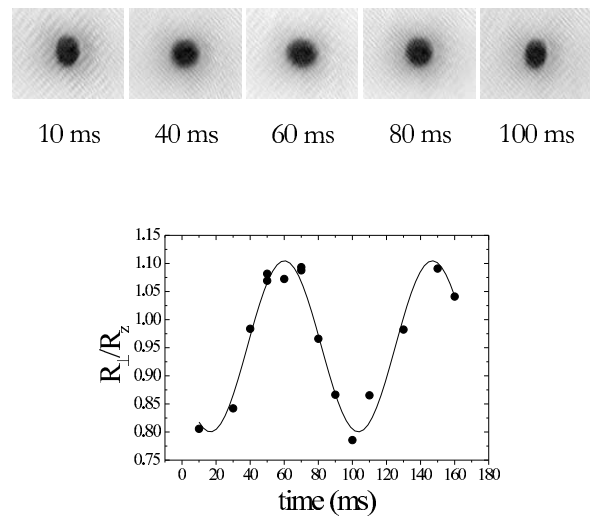}
\caption{In the upper part we show absorption images taken after
exciting the quadrupole mode of a condensate in the combined trap
with $V_{opt}=3.4$~$E_r$ and waiting different evolution times
(10,40,60,80 and 100~ms) before switching off the trap and letting
the cloud expand for 29~ms. In the lower part we show the
evolution of the Aspect Ratio ($R_{\perp}/R_z$) of the central
interference peak obtained from the absorption images together
with a sinusoidal fit to extract the frequency of the mode.}
\label{foto}
\end{center}
\end{figure}
A typical series of data is represented in Fig.~\ref{foto} where,
in the upper part, we show images of the expanded condensate taken
at different times after the excitation procedure and in the lower
part we report the measured aspect ratio together with the
sinusoidal fit. We repeated the same procedure for various
intensities of the laser light creating the optical lattice. Even
if the data were taken at finite temperature, we do not observe
any significant damping rate in the time scale of our experiment
where we follow the oscillation for $\sim 200$~ms. This is
consistent with previous results for the quadrupole mode in pure
magnetic traps \cite{MITT,noi}.

\begin{figure}
\begin{center}
\includegraphics[width=8cm]{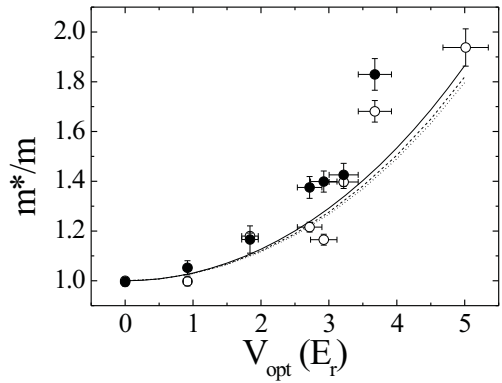}
\caption{Effective mass values extracted from the dipole mode
frequency (open circle) and from the quadrupole mode frequency
(closed circle) as a function of the optical lattice height. The
continuous line represent the theoretical curve from
\cite{kraemer} obtained neglecting the role of the mean field
interactions, while dashed and dotted lines corresponds to the
values obtained in \cite{massignan} numerically solving the
Gross-Pitaevskii equation and evaluating the effective mass from
the quadrupole and the dipole mode frequencies.} \label{massa}
\end{center}
\end{figure}

From Eqs.~(\ref{frequenze}) $\nu_D$ and $\nu_Q$ are expected to
scale in the same way with the optical potential depth.
\begin{figure}
\begin{center}
\includegraphics[width=8cm]{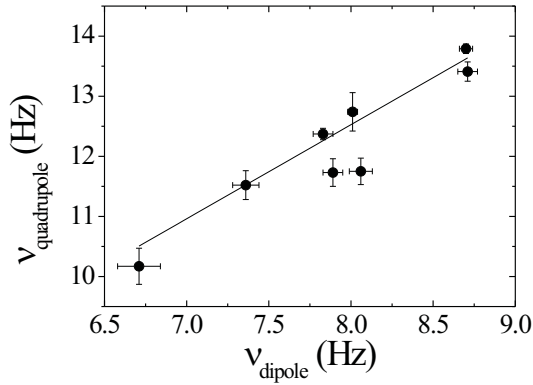}
\caption{Frequency of the quadrupole mode of a condensate trapped
in the combined potential (harmonic magnetic trap + 1D optical
lattice) as a function of the dipole mode frequency measured for
different values of the optical lattice depth from 0~$E_r$ to
4.1~$E_r$. The frequencies of both the modes, characterized by a
dynamics along the optical lattice, show a marked dependence from
the optical potential depth. The line represent a linear fit with
a slope of $1.57\pm0.01$.} \label{quadrupole}
\end{center}
\end{figure}
In Fig.~\ref{quadrupole} we report the quadrupole mode frequency
as a function of the dipole mode frequency varying the optical
lattice height up to 4.1~$E_r$. Both the dipole and the quadrupole
frequency exhibit a strong dependence on the lattice potential (we
observed a variation of $\sim 30\%$) demonstrating the marked
effect of the optical lattice on these modes. Furthermore, from a
linear fit of the data shown in Fig.~\ref{quadrupole} we obtain a
slope of $1.57\pm0.01$ in very good agreement with the theoretical
prediction of $\sqrt{5/2}=1.58$. Using Eqs.~(\ref{frequenze}),
from our data we can also extract the value of the effective mass
$m^*$ as a function of $V_{opt}$. The results obtained from both
the dipole mode and the quadrupole mode frequencies are reported
in Fig.~\ref{massa}, together with the theoretical predictions
reported in \cite{kraemer} (continuous line). Even if this
theoretical curves have been obtained neglecting the mean field
interaction and the magnetic confinement, the agreement with our
data is very good. In fact, in the regime of $V_{opt}$ explored in
our experiment, the effect of interactions is negligible
\cite{nota} as also confirmed by the direct solution of the
Gross-Pitaevskii equation (dashed and dotted line in
Fig.~\ref{massa}) \cite{massignan}. It would be interesting to
investigate also the regime of higher optical lattice depth where
the effective mass grows exponentially. On the other hand,
accessing this regime without entering instability regions seems
to be very difficult \cite{instabilita}.

\begin{figure}
\begin{center}
\includegraphics[width=8cm]{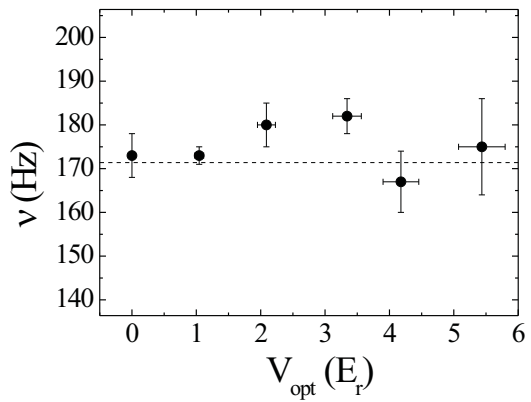}
\caption{Frequency of the transverse breathing mode of the
condensate in the combined trap as a function of the optical
lattice depth $V_{opt}$. The dashed line corresponds to the
expected value 2$\nu_{\perp}$.} \label{breathing}
\end{center}
\end{figure}

{\it Transverse Breathing Mode.} In a different series of
experiments we also excite the transverse breathing mode
modulating the bias field at a frequency close to 2~$\nu_{\perp}$.
In this case the excitation procedure takes only $\sim 30$~ms and
no evident heating of the condensate is observed (from the
condensed fraction of the cloud we can estimate a temperature
$T<0.8$~$T_c$). We repeat this measurement for different optical
lattice depths and the results are summarized in
Fig.~\ref{breathing}. The experimental data are in agreement with
the expected value 2$\nu_{\perp}$ and no dependence on the optical
potential height is observed. This confirms the prediction that
the dynamical behaviour of the condensate along the radial
direction (perpendicular to the lattice axis) is not affected by
the presence of the optical potential.

In conclusion, we have investigated the quadrupole and transverse
breathing modes of a harmonically trapped condensate in the
presence of a 1D optical lattice along the axial direction. The
frequency of the quadrupole mode, mainly characterized by an axial
oscillation of the cloud shape, shows an evident dependence on the
lattice height in agreement with the renormalized mass theory. On
the contrary, the transverse breathing mode, which in our geometry
predominantly occurs perpendicularly to the lattice axis, does not
exhibit any dependence on the lattice intensity. Our measurements
demonstrate that the transport properties of a trapped BEC in the
presence of a periodic potential can be described generalizing the
hydrodynamic equations of superfluids for a weakly interacting
Bose gas. From the measured frequency of the quadrupole and dipole
modes we extracted a value for the effective mass that is in very
good agreement with the predictions obtained even neglecting in
the calculation the effect of interactions.

We acknowledge S.~Stringari, M.Kr\"{a}mer and M.~Modugno for
fruitful discussions, theoretical contribution and the critical
reading of the manuscript. This work has been supported by the EU
under Contracts No. HPRI-CT 1999-00111 and No. HPRN-CT-2000-00125,
by the MURST through the PRIN 1999 and PRIN 2000 Initiatives and
by the INFM Progetto di Ricerca Avanzata ``Photon Matter''.

\end{document}